\documentclass[%
aip,
rsi,
superscriptaddress,
amsmath,amssymb,
reprint,%
]{revtex4-1}

\usepackage{physics} 
\usepackage{siunitx} 
\usepackage{graphicx}
\usepackage{dcolumn}
\usepackage{bm, float}
\usepackage{upgreek}
\usepackage[utf8]{inputenc}
\usepackage[T1]{fontenc}
\usepackage{mathptmx}
\usepackage{xcolor}
\usepackage{textcomp}
\usepackage{gensymb}
\usepackage{hyperref}
\usepackage{cleveref} 

\begin{document}

\preprint{AIP/123-QED}

\title{A mechanically stable and tunable cryogenic Fabry-Perot microcavity}
\author{Y. Fontana}
\thanks{These authors contributed equally to this work}
\affiliation{Department of Physics, University of Basel, Klingelbergstrasse 82, 4056 Basel, Switzerland}
\author{R. Zifkin}
\thanks{These authors contributed equally to this work}
\affiliation{Department of Physics, McGill University, 3600 Rue University, Montreal Quebec, Canada}
\author{E. Janitz}
\affiliation{Department of Physics, ETH Z\"{u}rich, Otto-Stern-Weg 1, 8093 Z\"{u}rich, Switzerland}
\author{C. D. Rodr$\text{\'i}$guez Rosenblueth}
\affiliation{Department of Physics, McGill University, 3600 Rue University, Montreal Quebec, Canada}
\author{L. Childress}
\email[corresponding author: ]{lilian.childress@mcgill.ca}
\affiliation{Department of Physics, McGill University, 3600 Rue University, Montreal Quebec, Canada}

\date{\today}

\begin{abstract}
High-finesse, open-geometry microcavities have recently emerged as a versatile tool for enhancing interactions between photons and material systems, with a range of applications in quantum optics and quantum information science. However, mechanical vibrations pose a considerable challenge to their operation within a closed-cycle cryostat, particularly when spatial tunability and free-space optical access are required. Here, we present the design and characterization of a system that can achieve ${\sim}16$ pm-rms passive mechanical stability between two high-finesse mirrors while permitting both three-dimensional positioning of the cavity mode and free-space confocal imaging. The design relies on two cascaded vibration isolation stages connected by leaf springs that decouple axial and lateral motion, and incorporates tuned-mass and magnetic damping. Furthermore, we present a technique for quantifying cavity length displacements similar to or larger than the cavity linewidth, allowing in-situ measurement of vibrations with and without active feedback. Our results facilitate operation of a tunable, high-finesse cavity within a closed-cycle cryostat, representing an enabling technology for cavity coupling to a variety of solid-state systems. 

\end{abstract}

\maketitle

\section{Introduction}

Fabry-Perot microcavities offer an opportunity to engineer light-matter interactions with exquisite in-situ control over the resonance frequency and location of the cavity mode. 
Such tunability enables optimization of the cavity spectral and spatial overlap with a quantum emitter,~\cite{liTunableOpenAccessMicrocavities2019, tommBrightFastSource2021} and facilitates scanned imaging techniques.~\cite{maderScanningCavityMicroscope2015} 
Furthermore, open-geometry cavities can attain high quality factor to mode volume ratios by combining state-of-the-art mirror coatings with novel small-radius-of-curvature mirror substrates,~\cite{trichetTopographicControlOpenaccess2015, najerFabricationMirrorTemplates2017} thereby strongly enhancing interactions between cavity-confined photons and quantum emitters. 
Moreover, by separating the optical resonator from the material system contained within, open microcavities can minimize fabrication-induced degradation of solid-state emitters.~\cite{rufOpticallyCoherentNitrogenVacancy2019}
As an example application, an open microcavity containing a nitrogen-vacancy (NV) center in a diamond membrane has been proposed as a mechanism to speed up entanglement distribution in quantum networks.~\cite{rozpedekNeartermQuantumrepeaterExperiments2019}

For the most advanced applications, operation at cryogenic temperatures is essential to reduce or eliminate the thermal phonon broadening present in solid-state systems. However, cryostat noise makes it challenging to achieve the mechanical stability required by high-finesse cavities while also maintaining the ability to position the cavity mode in three dimensions. Ideally, residual vibrations between the two cavity mirrors should be significantly less than the linewidth in length of the cavity resonance $\Delta L = \lambda/(2 \mathcal{F})$, where $\lambda$ is the resonant wavelength and $\mathcal{F}$ is the finesse; $\Delta L \sim 30$ pm for $\lambda = 600$ nm and $\mathcal{F} = 10,000$. For tunable systems based on wet cryostats, microcavity stability as low as 4.3 pm-rms~\cite{greuterSmallModeVolume2014} has been reported in a nitrogen-shield-free bath cryostat with the entire dewar acoustically shielded and resting on an actively vibrationally isolated platform. Cryostat noise poses an even greater challenge in closed-cycle coolers, where the compressor and other moving parts introduce a strong mechanical drive. In such systems, sub-pm-rms passive stability has only been achieved for cavities without 3D spatial tunability;~\cite{merkelCoherentPurcellEnhancedEmission2020} among systems incorporating a 3D positioner, passive stability of ${\sim}50$ pm-rms~\cite{rufResonantExcitationPurcell2021} and stability with active cavity locking of ${\sim}20$ pm-rms~\cite{casaboneDynamicControlPurcell2020} have been reported during quiet periods synchronized to the cryocooler cycle. 

Here, we present a cryogenic, widely tunable microcavity platform that achieves ${\sim} 16$ pm-rms passive mechanical stability during cryocooler quiet periods. This performance is enabled by combining a rigid positioner with a home-built vibration isolation platform (VIP), which are both mounted within a commercially available table-top closed-cycle cryostat. In addition to isolating the cavity from noise along its axis, the VIP is designed to permit confocal imaging of the system via free-space optics with sub-micron resolution, requiring lateral rigidity and decoupling of axial motion. We characterize the vibrations in our system using a technique that combines cavity transmission measurements with a Pound-Drever-Hall error signal; \cite{blackIntroductionPoundDreverHallLaser2001} this approach makes it possible to characterize relatively large displacements in-situ with a high-finesse cavity. We also consider active cavity locking at ambient and cryogenic temperatures, and discuss implications for future improvements. 

It is worth noting that the combination of inter-mirror stability and tunability required by microcavity applications is similar to the constraints on tip-sample distance in advanced scanning probe microscopy (SPM) experiments. Indeed, closed-cycle cryogenic SPM systems with stability at the picometer level have been achieved.~\cite{hackleyHighstabilityCryogenicScanning2014, zhangCryogenfreeLowTemperature2016, chaudharyLowNoiseCryogenfree2021} However, they are not compatible with free-space confocal imaging, as the SPM platform is typically suspended from highly compliant springs. The system and design principles presented here may thus also prove relevant to cryogenic SPM applications where precision optical imaging is needed.~\cite{pelliccioneScannedProbeImaging2016} 


\section{Design of the apparatus}                 
\begin{figure}
    \centering
    \includegraphics{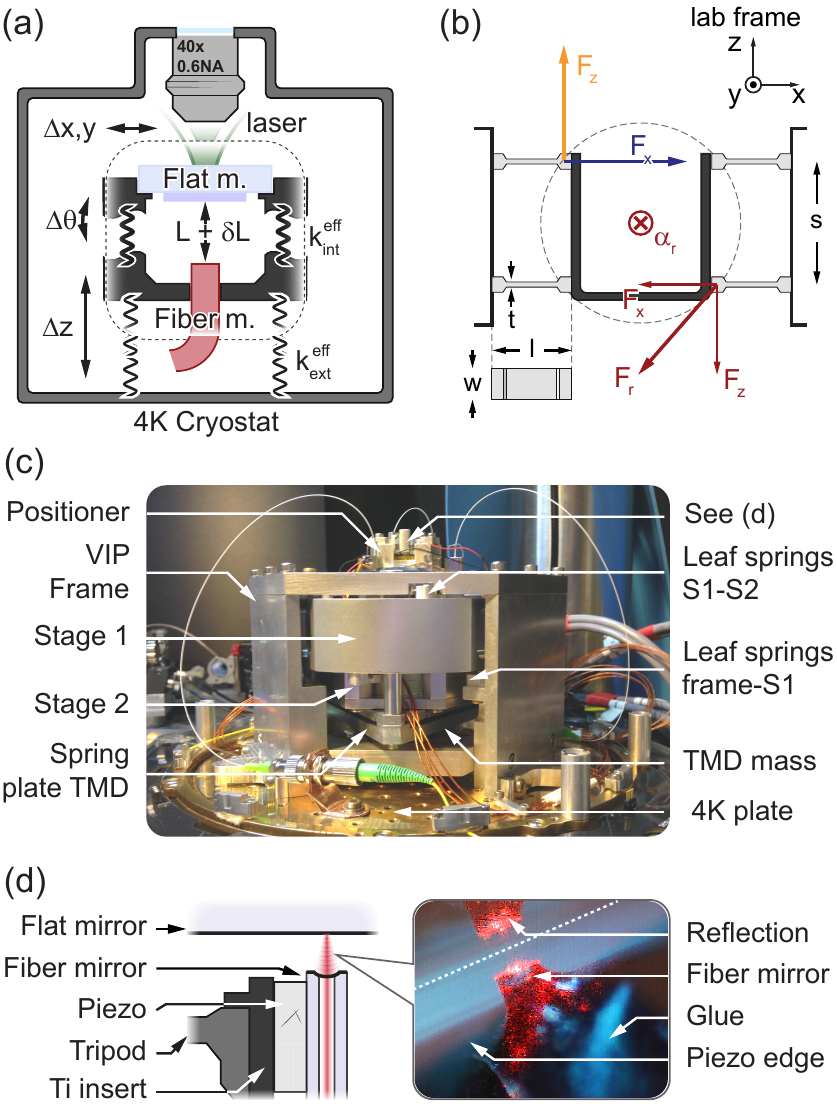}
     \caption{VIP design considerations. a) Representation of the cryogenic scanning fiber cavity. The axial displacement ($\delta L$) is sensitive to external vibrations ($\ll$ nm). The additional external degrees of freedom of the positioner (where $\Delta z$ is the axial translation, $\Delta x,y$ is the in-plane translation and $\Delta \theta$ is the tilting motion) can impede external access e.g. by a diffraction-limited laser beam. b) Concept of the dual-plane leaf-spring arrangement. Arrows depict a representative axial force (orange), transverse force (blue) and rotational forces and torque (red). c) Picture of the VIP and positioner supporting the cavity mounted on the 4K plate of the cryostat. d) Schematic of the fiber, shear piezo and titanium insert (left) and picture of the glued fiber (right).}
    \label{fig1}
\end{figure}

\begin{figure*}
    \centering
    \includegraphics{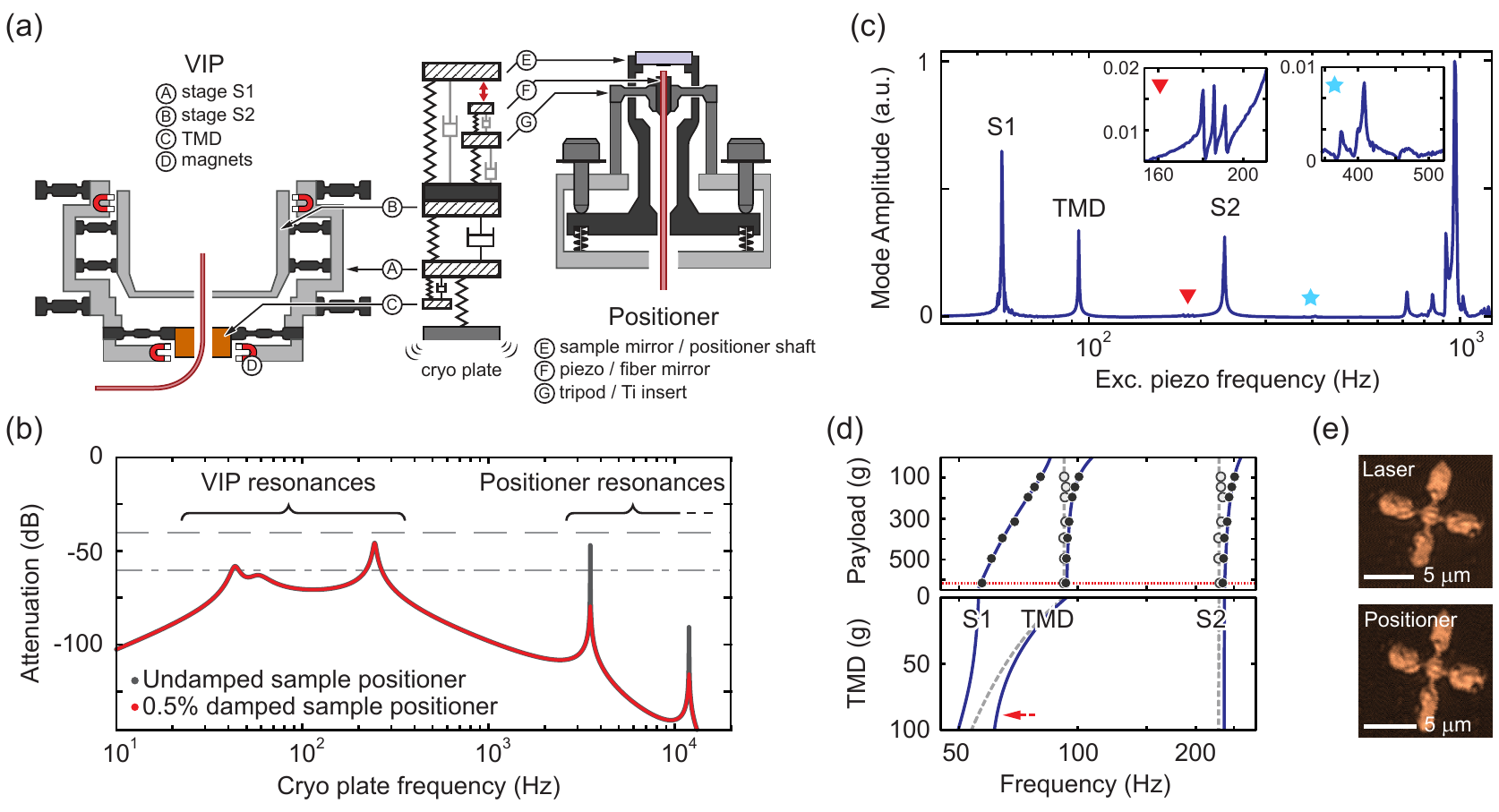}
    \caption{VIP modelling and characterization: a) One-dimensional damped mass-and-spring model for the passively stabilized cavity system, comprising the VIP (left) and the customized positioner (right), where the cryo plate drives the whole assembly. The fiber-sample spacing (cavity length) is indicated by a red arrow. b) Calculated inter-mirror motional amplitude attenuation relative to the cryo-plate amplitude using the model in (a) with a TMD tuning-load of 87 g and a payload of 615 g. The dashed (dotted-dashed) line marks attenuation by a factor $10^2$ ($10^3$). c) Sample mirror axial motion measured with respect to the lab frame upon excitation by a piezo located on S1, without any tuning-load and with a payload of $615$g (weight of the positioner). The main VIP axial modes (corresponding to S1, S2 and the out-of-tune TMD) are labelled. Insets depict residual motion attributed to rocking of S1 and S2. The high-frequency peaks near 1 kHz are assigned to localized, higher-order modes or the positioner internal resonances. d) Frequencies of the three main VIP resonances (solid gray circles) and two anti-resonances (empty black circles) measured while changing the VIP payload without tuning-load (top). The  red dotted line indicates the payload corresponding to the positioner. Blue solid and gray dashed lines are fits to a model comprising three masses and springs. The fit parameters are used to extrapolate the necessary tuning-load of 90 g (bottom, optimal value indicated by a dashed red arrow). e) Reflection confocal microscopy images of markers on the sample taken by scanning the laser beam (top) or the positioner piezos (bottom) during cryostat operation.}
    \label{fig2}
\end{figure*}
Ultimately, our goal is to reduce the effect of external vibrations on the relative motion between cavity mirrors, which in our setup are comprised of a small radius-of-curvature mirror on the tip of an optical fiber and a macroscopic flat mirror to which a sample can be bonded. The mechanical assembly binding the two mirrors effectively implements a mechanical high-pass filter, protecting the system against low-frequency components of the vibration spectrum. The cutoff frequency of this filter is determined by the mechanical resonance frequencies of the positioner, which are maximized by reducing the effective mass and increasing the rigidity. A stiff and massive positioner also makes it possible to rapidly vary the inter-mirror spacing without inducing significant motion of the surrounding elements, which is crucial for active cavity stabilization.

Unfortunately, mechanical resonance frequencies over \SI{1}{kHz} are hard to achieve due to the complexity of positioners, design considerations and ultimately material properties, setting an upper frequency limit to noise-rejection. However, it is possible to prevent high-frequency noise from reaching the positioner by adding an isolation stage.~\cite{pohlDesignCriteriaScanning1986, okanoVibrationIsolationScanning1987, parkTheoriesFeedbackVibration1987} 
In its simplest form, an isolation stage can be realized by suspending the apparatus with springs, effectively forming a harmonic oscillator acting as a low-pass filter. Softer springs or a heavier suspended mass lead to a lower cutoff frequency and therefore more aggressive filtering of the vibration spectrum. Various types of suspension systems (coil springs, levitation, multiple stages) and damping strategies can be used depending on the particular application and the conditions in which the instrument operates (low temperature, vacuum etc.).~\cite{binnigTunnelingControllableVacuum1982, astDesignCriteriaScanning2008, okanoVibrationIsolationScanning1987, zhangUltrastableSiliconCavity2017, dewitVibrationIsolationHigh2019}
For the most sensitive instruments, a combination of a rigid assembly and an isolation platform is necessary.

Our scanning cavity system is subject to a number of design constraints that are frequently encountered in modern setups: it operates in high vacuum (${\sim}$\SI{1e-6}{mbar}), at cryogenic temperature (cryo plate temperature of \SI{4}{K}), and in a table-top, closed-loop cryostat (a first-generation Montana Instruments Nanoscale Workstation). Our particular experimental application involves introducing an optically active, solid-state sample (a diamond membrane containing single emitters) into the microcavity by bonding it to the flat mirror. In addition to maintaining a stable inter-mirror distance, we also wish to image the sample at multiple wavelengths by exciting and collecting fluorescence with a high-numerical-aperture objective mounted to the frame of the cryostat. Such external optical access precludes the use of an arbitrarily compliant isolation platform as it must be stiff enough to enable confocal imaging with sub-micron resolution. A generic depiction of such a setup is shown in Fig. \ref{fig1}; while the effective internal rigidity of the positioner $\text{k}^{\text{eff}}_{\text{int}}$ should be as high as possible for all degrees of freedom, the constraints on the spring constant $\text{k}_{\text{ext}}^{\text{eff}}$ of the vibration isolation platform can be summarized by three conditions:
\begin{enumerate}
\item  $\text{k}_{\text{ext}}^{\text{eff}}$ in the transverse ($x,y$) directions should be as high as possible ($\Delta x,y$, $\Delta\theta <$ \SI{1}{\micro\meter}),
\item  $\text{k}_{\text{ext}}^{\text{eff}}$ along the axial ($z$) direction should be high enough to maintain our sample within the focal range ($\Delta z {\sim}$ \SI{1}{\micro\meter}),
\item  $\text{k}_{\text{ext}}^{\text{eff}}$ along the axial direction should be as low as possible to minimize inter-mirror vibrations ($\delta L {\sim}$ \SI{10}{\pico\meter}-rms).
\end{enumerate}

Conditions 1 and 3 require that the platform is engineered such that axial motion is effectively decoupled from other degrees of freedom. Conditions 2 and 3 indicate that a compromise must be made between vibration attenuation and relative motion between the objective and the sample in the axial direction.
These conditions rule out the use of coil springs for suspension, as their marginal transverse rigidity would lead to pendulum swing and rocking motion (corresponding to large $\Delta xy$ and $\Delta \theta$ amplitudes). Furthermore, spring extension can lead to long, cumbersome assemblies, for which cooling the system without an exchange gas is impractical. 
One solution that fulfills all three conditions employs multiple stacked plates spaced by elastomeric joints, an approach commonly used in scanning probe apparatus.~\cite{okanoVibrationIsolationScanning1987, olivaLowHighfrequencyVibration1998} 
In such designs, the reduced axial compliance is offset by the multiplication of the stages: the resonant frequency is higher, but the roll-off is steeper. However, operating at cryogenic temperature prevents the efficient use of elastomers as compliant, damping elements.

In order to circumvent these challenges, we propose a VIP based on leaf springs (see supplementary material~\cite{sup-cad} for details), which can be designed to attain a desired spring constant for bending while maintaining very high rigidity for all other deformation. 
A two-dimensional schematic of our “dual-plane” arrangement can be seen in Fig.~\ref{fig1}b, where a central platform (dark gray) is held by four leaf springs tethered to a rigid base. 
The springs are arranged on two horizontal planes separated by distance $s$. 
The motivation for this design can be understood by considering how the central platform displaces in response to axial ($\Delta z$, orange), transverse ($\Delta xy$, purple) and rotational ($\Delta \theta$, red) forces. 
The use of leaf springs allows us to separately engineer the response to transverse and axial forces.
Using rectangular homogeneous beams with thickness $t$, width $w$ and length $L$ as a toy model for the leaf springs, the ratios of resonant frequencies along the axial (thickness) direction $z$ and the transverse directions $x$ and $y$ are simply
\begin{equation}
    \frac{\omega_z}{\omega_x} = 2\sqrt{2}\frac{t}{L} \mathrm{\quad and \quad } \frac{\omega_z}{\omega_y} = \frac{t}{w},
\end{equation}
allowing us to tune the axial compliance while maintaining transverse rigidity.
In the case of a torque applied to the platform, a force acts on the springs with both $F_{z}$ and $F_{x,y}$ components (red arrows in Fig.~\ref{fig1}b). The exact projections of the forces along $z$ and $x,y$ will depend on the geometry and distribution of the load on the platform; for the simple case of a platform with center of rotation halfway between the two planes of springs, we find that the contribution of $F_{x,y}$ depends linearly on the inter-plane distance $s$. Consequently, for large $s$, the load transfer to axial bending is reduced, corresponding to an increase in torsional rigidity. By implementing such a dual-plane leaf-spring arrangement with a large inter-plane spacing $s$, axial motion can be effectively decoupled from all other modes.

We focused our design primarily on the axial modes since vibration isolation along $z$ is the most stringent requirement. We tackled the contradictory conditions $2$ and $3$ by nesting two platforms, hereby referred to as S1 and S2. Such a cascaded setup allows us to maintain relatively high S1 (${\sim}$\SI{50}{Hz}) and S2 (${\sim}$\SI{230}{Hz}) resonant frequencies while benefiting from increased roll-off at higher frequencies.~\cite{okanoVibrationIsolationScanning1987}
Given the target resonant frequencies of S1 and S2, and the approximate load, we designed the springs to have the necessary rigidity by using basic beam theory estimates followed by a more complete calculation using a finite element analysis software\footnote{Ansys\textregistered~Workbench, 17$^\mathrm{th}$ edition, Ansys Inc.}. We then fine-tuned the geometrical parameters to obtain resonant modes along $z$ at the target frequencies while maximizing the frequency of all other modes (rocking, twisting and lateral translation).

Another essential element of the isolation design is damping, as it suppresses large-amplitude motion at the platform resonant frequencies. Unfortunately, operating at cryogenic temperature precludes the use of elastomers as damping elements, while dashpot-type damping is incompatible with vacuum. Since adding significant damping directly to the S1 platform would decrease the roll-off after resonance, we employ eddy-current magnet braking~\cite{sodanoImprovedConceptModel2006} applied on a tuned-mass damper (TMD) anchored to S1 with a spring plate.~\cite{frahmDeviceDampingVibrations1909} The spring plate in itself contributes to the effective mass of the TMD mechanical mode. This mass is further increased by attaching a tuning-load made of copper to the spring plate. The tuning-load mass is chosen so that the first axial mode of the TMD hybridizes with the S1 mode; upon excitation, the energy is transferred back and forth between the platform and the TMD, which in turn is damped by a set of six samarium cobalt (SmCo) magnets arranged around the tuning-load to minimize the stray magnetic field at the sample location. Due to spatial constraints, S2 is damped directly by three pairs of SmCo magnets arranged with the same consideration, resulting in a reduction in roll-off that comes as an engineering compromise. We obtain further damping by constructing all other components from aluminum, which has a comparatively high residual damping parameter among different metals at low temperature. This material choice carries other considerations as well. Aluminum has a ratio of Young's modulus to weight roughly equivalent to stainless steel and titanium yet it is easily sourced and machined with modern tools, ensuring a finished product with tight tolerances. Furthermore, while aluminum has the disadvantage of high thermal contraction, in principle it provides excellent thermal conductivity; our observed sample temperature of approximately 11K (cryo plate nominal temperature of 4K) was likely limited by imperfect thermal contacts, which could be improved by application of indium or other thermally conductive material. 

The assembled VIP supports a customized JPE (Janssen Precision Engineering) CPSHR3-S-COE sample positioner, which relies on a proprietary design to achieve long (${\sim}$\si{\milli\meter}) travel range in closed-loop mode while maintaining high rigidity with a specified unloaded axial (transverse) resonant frequency of \SI{3.7}{kHz} (\SI{1.5}{kHz}).\footnote{JPE B.V., \url{https://www.jpe-innovations.com/}} 
The positioner is mounted in the VIP on S2 (see Fig.~\ref{fig1}c). Within the positioner, the mounting of the fiber and sample mirrors is also important to the mechanical properties of the system; the fiber is supported directly on the rigid housing of the positioner via a tripod, while the sample mirror can be moved. 
The fiber tip is facing upward and is glued to a high-frequency (${\sim}$\SI{1}{MHz} unloaded) shear piezo which allows for fine adjustment of the fiber position along $z$, thereby tuning the cavity resonance (see Fig. \ref{fig1}d). 
The piezo is mounted on a 3D-printed titanium insert, which is tightly nested in the tripod mount and held in position by a screw. 
To maximize the rigidity of the fiber mirror mount, we designed the tripod to have a high frequency (\SI{15}{kHz}) for its lowest-order drum-like mechanical mode, and glued the fiber with minimal overhang to suppress cantilever oscillations.  
A close-up view of the assembled cavity is displayed in Fig. \ref{fig1}d (right), showing  the fiber mirror as well as its reflection in the flat sample mirror.

As discussed above, our leaf-spring-based, dual-plane arrangement essentially decouples the $z$ modes from other mechanical modes, allowing us to model our stage and positioner as a 1D system of coupled harmonic oscillators. Figure~\ref{fig2}a shows representative schematics of the VIP (left) and the JPE positioner (right) as well as the equivalent model, which can be used to calculate the amplitude of cavity length excursions $\delta L$ (red arrow) as a function of the amplitude of cryo plate displacement $A_{\mathrm{cryo}}$ (the main source of mechanical noise) over a range of frequencies.  The results, expressed as $20 \log{(\delta L/A_\mathrm{cryo})}$, can be seen in Fig.~\ref{fig2}b for two different values of positioner damping, where the dashed (dotted-dashed) line indicates an attenuation of $\delta L/A_\mathrm{cryo}=10^{-2}$ ($10^{-3}$).  
The resonances of the VIP (positioner) are clearly visible at low (high) frequencies. Larger frequency splitting between VIP and positioner modes leads to larger attenuation over the full spectrum; at frequencies below the positioner resonances, the two mirrors move together, while at frequencies above the VIP resonances, the amplitude of transmitted oscillations are strongly suppressed.

To characterize the VIP, we measured the resonance frequency of its modes using an unbalanced Michelson interferometer under ambient conditions. One arm of the interferometer served as a reference while the mirror of the second arm was mounted on the VIP top platform (S2). Measurements were performed using a tunable 637 nm laser with its wavelength adjusted to maximize the sensitivity of the resulting photodiode signal.  The laser frequency was then locked to the diode signal, resulting in a frequency detuning that was directly proportional to the axial motion of S2. A piezo shaker, mounted off-center on the bottom platform S1, provided a local source of excitation. The shaker was driven by a lock-in amplifier, which was also used to demodulate the measured motion of S2 and the sample mirror, allowing us to strongly drive the stage and measure its response with excellent signal to noise.

The VIP spectral response comprises three main peaks in the \SIrange{50}{320}{Hz} range (see Fig.~\ref{fig2}c) that correspond to modes stemming from the TMD, S1 and S2. The insets of Fig.~\ref{fig2}c provide a close-up of low-amplitude modes around \SI{200}{Hz} and \SI{400}{Hz} related to rocking motion of S1 and S2. These resonance triplets mirror the three-fold symmetry of the stage, slightly broken by the off-center drive and load imbalance. At higher frequencies, large-amplitude peaks start to appear that cannot be attributed with certainty but probably originate from local internal vibrational modes of S1 efficiently excited by the piezo or the positioner internal resonances. Provided that these modes are localized, their role in transferring vibrations from the cryo plate should be minimal.
The spectrum in Fig.~\ref{fig2}c was acquired with a full payload (the positioner) but without any TMD tuning-load; in order to find the correct tuning-load mass leading to S1 and TMD mode hybridization, we varied the payload from \SI{100}{g} to \SI{615}{g} and recorded the position of the three main resonances and two anti-resonances (Fig.~\ref{fig2}d, top). We then fit a three-harmonic-oscillator model to retrieve the effective masses and spring constants, and infer a best value for the tuning-load ${\sim}$\SI{90}{g} for the full payload (Fig.~\ref{fig2}d, bottom).

The low amplitudes and relatively high frequencies of the rocking modes detected in Fig.~\ref{fig2}c indicate a high rocking and in-plane mode rigidity, necessary to fulfill condition 1 for imaging. With the full system mounted in the cryostat, we verified that we can image a micron-sized alignment marker by scanning either the laser beam or the fine piezos of the JPE (Fig.~\ref{fig2}e) during cryostat operation. Both images clearly resolve sub-micron features. The distortion in the piezo scan is due to the piezo hysteresis and the particular actuation mode of the positioner. Importantly, the resolution is maintained despite driving the positioner piezo in steps and at a high rate (several hundred Hz).

\section{Cavity length calibration}
\begin{figure}[htb]
    \centering
    \includegraphics{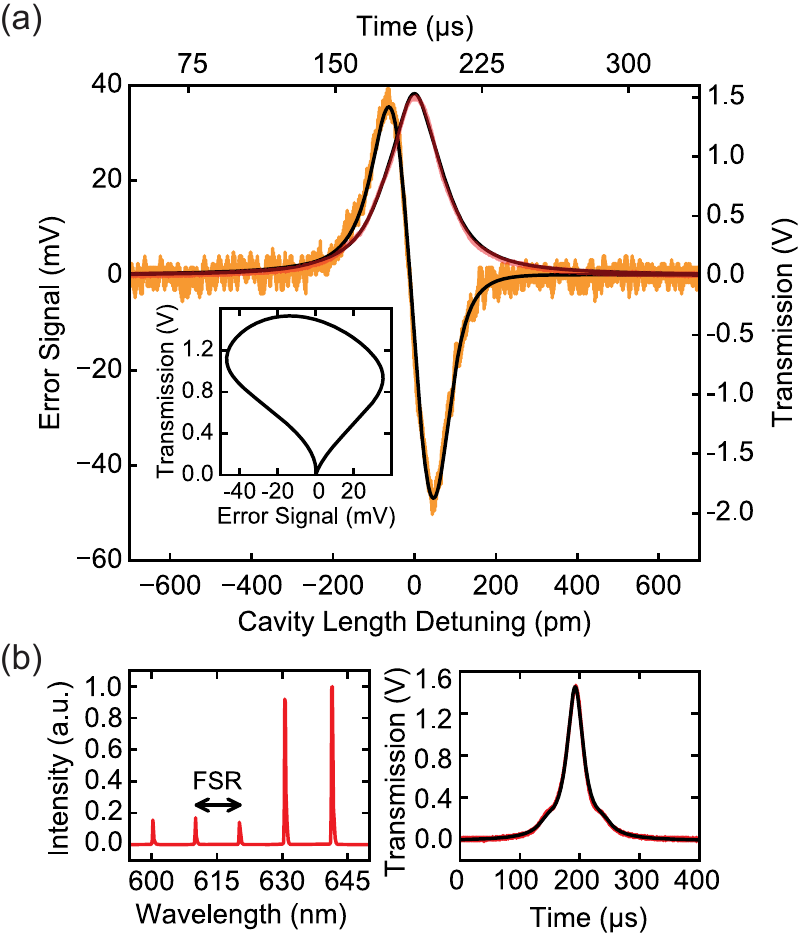}
    \caption{Cavity length extraction.  (a) Simultaneously acquired transmission and error signals as a function of cavity length with joint analytic fit overlaid. Inset shows the parametric plot of the fit. (b) Cavity broadband transmission spectrum (left), used to extract cavity length from the FSR ($L = 18.7 \pm 0.7 $ µm), and transmission of a 652.6 nm laser with sidebands at 4.5 GHz (right), used to determine cavity linewidth ($\Delta \nu=3.04 \pm 0.25$ GHz).}
    \label{fig3}
\end{figure}
We quantify the performance of the vibration isolation system by characterizing the transmission and reflection properties of the cavity it supports.
Notably, to increase the dynamic range for measuring cavity length excursions when working with high-finesse cavities, we have developed a technique that combines measurements of cavity transmission and a Pound-Drever-Hall (PDH) error signal.~\cite{blackIntroductionPoundDreverHallLaser2001} 
The transmission is symmetric about resonance, whereas the PDH signal is antisymmetric (see Fig.~\ref{fig3}a), such that the combined measurement allows us to unambiguously determine the cavity length excursion even for fluctuations comparable to the full cavity linewidth. 
This approach is applicable both with the cavity free-running and when we use the PDH error signal to lock the cavity to a reference laser.

The first step in measuring cavity length excursions is to calibrate the cavity transmission and PDH error signal as a function of the cavity length. 
Essentially, we convert measured signals as a function of time (as we displace a cavity mirror), to signals as a function of length by fitting our data to functional forms incorporating the cavity linewidth in length $\Delta L$ (see Figs.~\ref{fig3}c,d). The appropriate analytical expressions are found by calculating transmission and PDH error signals for a cavity with a complex coupling coefficient,~\cite{janitzFabryPerotMicrocavityDiamondbased2015} and are given in the supplementary materials.\cite{sup-cad}
We determine the linewidth in length $\Delta L$ via the finesse $\mathcal{F} = \mathrm{FSR}/\Delta \nu$, which is the ratio of two quantities we can directly measure: the cavity free spectral range (FSR, measured by white light transmission see Fig.~\ref{fig3}b, left) and the cavity linewidth in frequency ($\Delta \nu$, obtained from the cavity transmission of a phase-modulated laser whose sideband spacing provides a frequency reference, see Fig.~\ref{fig3}b, right). Matching the fitted linewidth in time to the known linewidth in length calibrates the $x$-axis for the error signal and transmission. 


We can readily confirm that combined measurements of the transmission and PDH error signal provide a unique determination of cavity length displacement by verifying that they do not self-intersect when plotted parametrically (see inset to Fig.~\ref{fig3}a). In practice, our measurement noise leads to overlapping data near the origin for positive and negative excursions that are significantly larger than the cavity linewidth $\Delta L$. Therefore, to maximize the dynamic range of our displacement measurement, we work with a lowered cavity finesse $\mathcal{F}=2700 \pm 20$ by using a long cavity ($L=18.7 \pm 0.7\ \upmu$m) where clipping losses dominate,~\cite{benedikterTransversemodeCouplingDiffraction2015} resulting in a linewidth of $\Delta L = 124.1 \pm 1.0$ pm.

\begin{figure}[htb]
    \centering
    \includegraphics{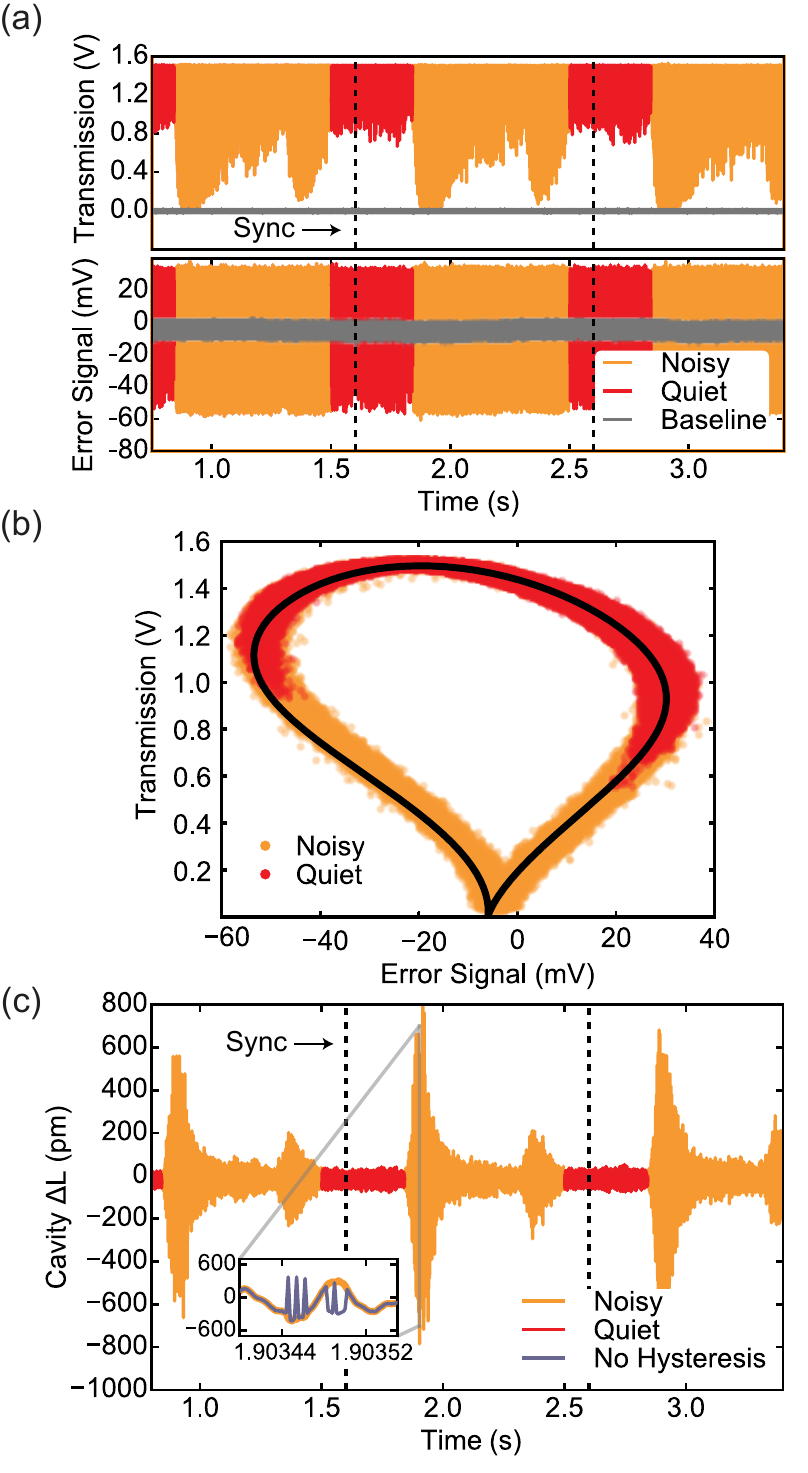}
    \caption{Unlocked cavity measurements. (a) Transmission and error signal as a function of time. The sync signal from the cryostat is indicated with a dotted vertical line and is used to distinguish between quiet periods (red), and the noisy periods (orange). Baseline (off-resonance) noise is also shown (gray). (b) Parametric plot of transmission and error signal from the data in (a), with quiet (red) and noisy (orange) data and the analytic fit is overlaid. (c) Cavity length excursion obtained assuming a 100 pm maximum jump in length between data points. Inset: comparison with (orange) and without (purple) the maximum jump restriction, taken during a noisy period.}
    \label{fig4}
\end{figure}

To acquire data on our cavity length fluctuations, we take noisy time traces of the error signal ${E_i \pm \sigma_E}$ and transmission ${T_i \pm \sigma_T}$, as shown in Figs.~\ref{fig4}a and b, where the gray data illustrates the measurement noise observed with the cavity off resonance.
Measurement noise creates some spread in the data around the ideal values, as illustrated in the parametric plot of Fig.~\ref{fig4}c. 
Here, the solid line shows the fits from Fig.~\ref{fig3}a, scaled and offset to best fit the data set (accounting for drifts in laser power and residual amplitude modulation in the error signal). 
Each point $(E(\delta L), T(\delta L))$ on the solid line thus corresponds to a known cavity length detuning $\delta L$. 
For a given measurement $(E_i\pm \sigma_E, T_i\pm \sigma_T)$, we infer the length detuning $\delta L_i$ by minimizing the noise-weighted distance ($r(\delta L_i)$) between the point $(E_i, T_i)$ and $(E(\delta L_i), T(\delta L_i))$, given by 
\begin{equation}
r(\delta L_i) = \sqrt{\left(\frac{E_i - E(\delta L_i)}{\sigma_E}\right)^2 + \left(\frac{T_i - T(\delta L_i)}{\sigma_T}\right)^2}.\label{noise_dist}
\end{equation}
For far-detuned cavities, both signals approach zero, so a sequence of measurements may be erroneously inferred to jump from far-positive to far-negative detuning. 
This is mitigated by limiting the range of $\delta L_i$ during the minimization procedure to $\delta L_{i-1}\pm\SI{100}{\pico\meter}$; the inset to Fig.~\ref{fig4}c illustrates the effect of this restriction.

\section{Results}
The data in Fig.~\ref{fig4} are acquired with the cavity free-running near resonance (unlocked), allowing us to evaluate the performance of our passive vibration isolation system. Even in the raw data, one can see that the cavity vibrations are strongly influenced by the cold head cycle of our cryostat. We therefore divide the data set into noisy (orange) and quiet (red) periods, which, in post-processing, are timed relative to a sync pulse provided by the cryostat. We note that the quiet period data is more reliable: as shown in Fig.~\ref{fig4}c, $(E,T)$ data acquired during the quiet period never approaches the origin of the parametric curve, staying far from the region where the curve intersects itself, and thus producing unambiguous measurements of cavity displacement. Consequently, within the quiet period we extract a reliable measurement of $16\pm1$ pm-rms, while the estimate of motion over the full cold head cycle ($54\pm4$ pm-rms) likely suffers from systematic errors. Here, the error bars are statistical errors extracted from the standard deviation of multiple measurements. The excellent passive stability achieved during the quiet region, which comprises ${\sim}34$\% of the complete cold-head cycle, creates an opportunity to perform measurements without the overhead of active cavity stabilization.


\begin{figure}[htp]
    \includegraphics{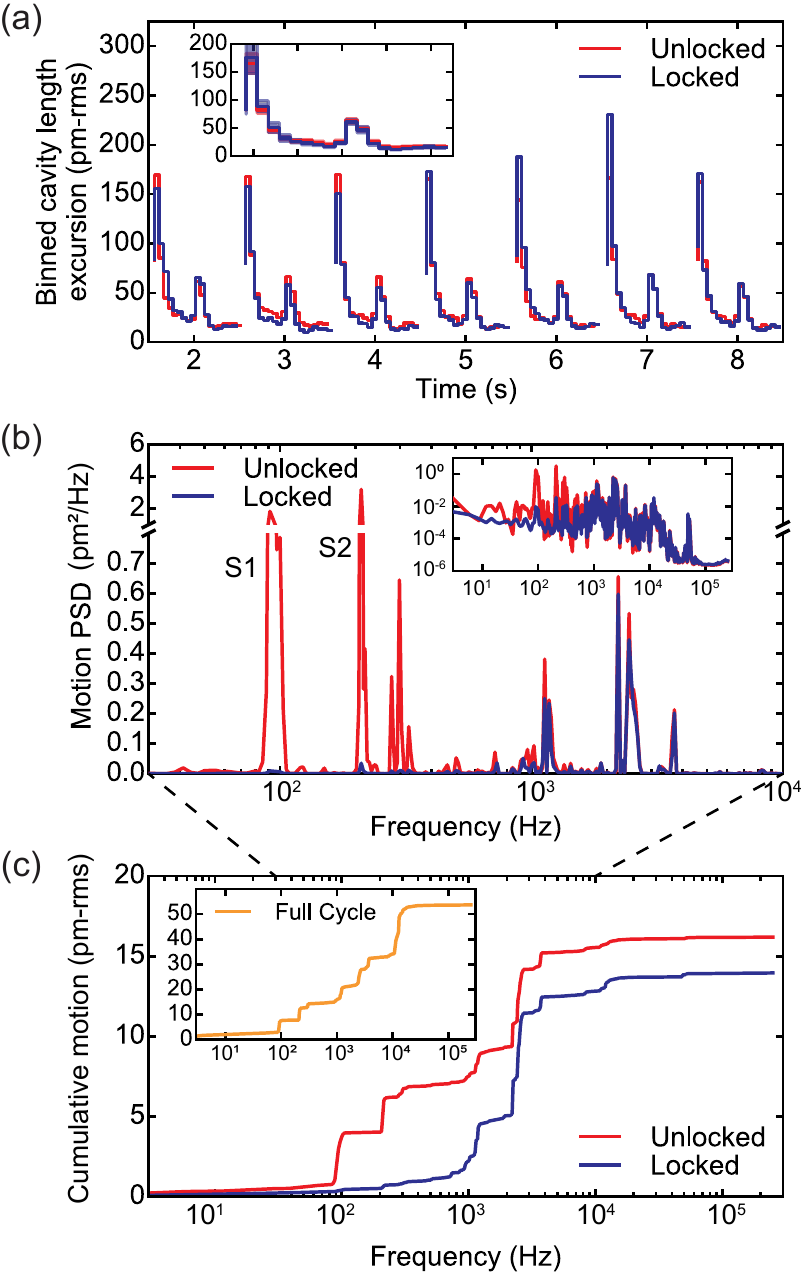}
    \caption{Locked cavity measurements. (a) The rms value of the cavity length excursions in 50 ms time bins for the locked (blue) and unlocked (red) cavity. Inset: Average rms value in a compressor cycle. The error bars denote the statistical uncertainty from averaging 9 cycles. 
    (b) PSD of length noise on a linear scale, with assignments of low-frequency peaks to VIP stage resonances (inset: log scale). (c) Cumulative cavity length excursion (rms) of the quiet period for the locked (blue) and unlocked (red) cavity. Inset: Full cycle unlocked (orange) cumulative cavity length excursion (rms).}
    \label{fig5}
\end{figure}

We can use the same measurement technique to characterize cavity vibrations while the cavity is actively stabilized. We lock the cavity to the laser frequency by feeding the PDH error signal back to the shear piezo on which the fiber mirror is mounted (Fig.~\ref{fig1}d). Figure~\ref{fig5}a shows the rms vibrations of the cavity with and without active stabilization, calculated in 50 ms intervals throughout the cold-head cycle, indicating that cavity locking does not substantially improve the performance of our system. 

Frequency-domain analysis provides greater insight into the performance of both passive and active stabilization. The power spectrum of vibrations during the quiet period is shown in Fig.~\ref{fig5}b for a free-running (red) and locked (blue) cavity. Active stabilization clearly improves rejection of low-frequency noise, up to approximately 1 kHz, but fails to suppress the dominant high-frequency noise. In fact, we see an increase in 1-3 kHz noise when active stabilization is enabled, as illustrated by the quiet period cumulative distribution in Fig.~\ref{fig5}c. Integrating the noise up to \SI{250}{\kilo\hertz}, we find an rms motion of $14.0 \pm 0.9$ pm-rms for the actively-stabilized cavity during the quiet periods, which is a marginal improvement over the $16\pm1$ pm-rms measured for the free-running cavity. 

Notably, this frequency-domain analysis also reveals resonant frequencies of the vibration isolation platform and the positioner. In particular, the low-frequency peaks S1 and S2 are shifted from their design values due to a combination of change in Young's modulus with temperature and thermal contraction, such that the TMD  becomes out-of-tune at low temperature. The absence of a visible TMD peak offers additional evidence that the TMD is not performing as designed. Improved modelling that takes into account shifts in material properties with temperature may offer a route to further improve the passive vibration isolation performance. 


\begin{figure}[htp]
    \includegraphics{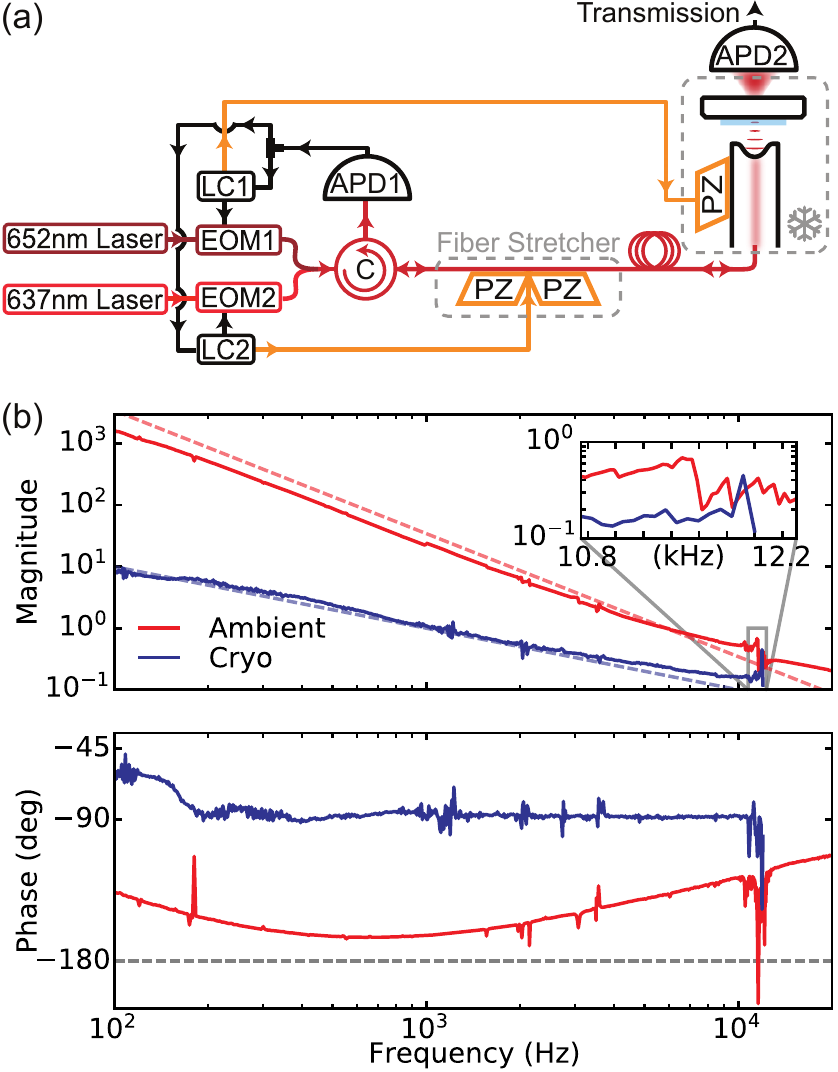}
    \caption{Cavity-locking characterization. (a) Experimental schematic: two lasers (652 and 637 nm) are phase modulated via electro-optic modulators (EOM1,2) at different frequencies; the beams are combined at the input of a circulator (C) and subsequently coupled to the microcavity through the fiber mirror. The reflected light is routed via the circulator to an avalanche photodiode (APD1) and the detector output is demodulated by two PDH circuits (LC1,2). The resulting error signals (shown in yellow) are fed back to two piezo-electric transducer (PZ) circuits: the first adjusts the position of the cavity mirror, and therefore the cavity length, and the second controls a stretcher for the cavity fiber. The light transmitted through the cavity is free-space coupled to a second detector (APD2). (b) Ambient (red) and low temperature (blue) circuit transfer functions. Blue (red) dashed lines indicate one (two) poles of roll off. The gray dashed line indicates a phase of -180\textdegree, around which the phase response is unwrapped. Inset: amplitude of the resonance that limits the locking bandwidth.}
    \label{fig6}
\end{figure}
The surprisingly poor performance of the active stabilization is rooted in several elements of our apparatus. 
One limitation arises from the finite dynamic range of our error signal, which approaches zero for displacements exceeding approximately the cavity linewidth. In the presence of large-amplitude disturbances, the feedback gain becomes nonlinear and, for extreme displacements, insufficient to maintain the lock. The dynamic range can be increased while maintaining the lock sensitivity by either decreasing cavity finesse (at the cost of higher laser power) or increasing the modulation frequency used for PDH locking (with diminishing effect once in the resolved-sideband limit). 

This finite dynamic range poses a even greater problem in the presence of fluctuations in the error-signal offset, which create a second zero-crossing point in the error signal. This marks the onset of positive feedback, pushing the cavity away from resonance.
In our system, such an offset arises from residual amplitude modulation (RAM) in the incident light, caused by a low-finesse interferometer formed between our fiber mirror and a weak reflection off the circulator, which converts phase modulation (introduced for PDH locking) into amplitude modulation with an efficiency that varies with interferometer length.
Even after optimizing our phase modulation frequency \cite{ehlersUseEtalonimmuneDistances2014}, small changes in the length of the optical fiber between our circulator and fiber mirror (due to variations in temperature and strain) induced significant fluctuations in the error signal offset. To maintain a robust cavity lock, it was therefore necessary to add a second, independent PDH feedback loop to actively stabilize the cavity fiber length (see Fig.~\ref{fig6}a) using a home-made fiber stretcher.
While this solution allowed us to maintain a cavity lock on the order of minutes, the slow variation in rms motion observed during the noisy periods in Fig.~\ref{fig5}a is likely an artifact of remaining weak RAM fluctuations.


The final and most critical challenge to active stabilization is the high frequency ${\sim}$kHz noise present in our system. 
We can characterize the frequency-dependent performance of our lock by its transfer function $G(f)$,~\cite{bechhoeferFeedbackPhysicistsTutorial2005} where cavity vibrations are suppressed by $|1+G(f)|$ when locked; Fig.~\ref{fig6}b shows the magnitude and phase of $G(f)$ for both ambient (red) and cryogenic (blue) operation. Prior to cooling down, we optimized our feedback system under ambient conditions, and found it was possible to achieve a ${\sim}$10 kHz lock bandwidth (defined as the lowest frequency where $|G(f)| = 1$) and nearly two poles of roll-off at low frequencies, leading to a factor of $40-50$ suppression in noise at 1 kHz and an approximately two orders of magnitude reduction in rms motion overall.

Unfortunately, the performance at cryogenic temperatures was markedly worse. While the ringing frequency (${\sim}$11 kHz) did not change significantly upon cool down, we were not able to achieve a stable lock with two poles of roll off at low frequencies or a bandwidth beyond 1 kHz without inducing ringing. These difficulties likely result from the change in material properties (most notably a reduction in material damping) at low temperature.
Indeed, the resonance near 11 kHz shows a large increase in amplitude (and decrease in width), which prevented us from further increasing the gain or introducing filters that decrease the phase near this frequency.
This resonance is likely associated with the tripod support for the fiber mirror, which could be improved through careful mechanical engineering. The relative insensitivity of the frequency upon cooldown, while surprising, could arise from the complicated interplay of thermal contraction, slippage and change in Young's modulus.
Such elusive details would justify a more complete characterization and modelling of the system resonances at low temperatures, which would allow improved filter design, bringing the cryogenic transfer function closer to the optimized ambient system, and potentially enabling 1-2 orders of magnitude improvement in system stability.

\section{Conclusion and outlook}
The design reported and characterized here enables a widely-tunable yet stable mechanical system within a table-top closed-cycle cryostat. Combining our home-built vibration isolation platform with a customized commercial positioner realizes a system with mm-scale tunability and $10\ \upmu$m-scale fine control that nevertheless demonstrates excellent passive stability (16 pm-rms) during quiet periods of the cold head cycle. Notably, our design exhibits this stability in an early-generation table-top closed-cycle cryostat that has its Gifford-McMahon cryocooler mounted on the optical table next to the sample chamber. Using a newer or heavily modified cryostat design that better decouples the closed-cycle cooler from the cold plate would lead to immediate improvements.~\cite{zhangCryogenfreeLowTemperature2016, chaudharyLowNoiseCryogenfree2021, zhangUltrastableSiliconCavity2017}
Moreover, the system would benefit from further optimization of the mechanical design, specifically the TMD at cryogenic temperature, as well as the structures of the tripod and fiber insert.
The tripod plays a crucial role for active feedback, and increasing both its mass and frequency would lead to higher locking bandwidth by shifting the limiting resonance frequency and diminishing its amplitude. Similarly, a more integrated insert mechanism would ensure that the insert itself does not introduce unwanted resonances in the system. 
Despite challenges that impede active locking, the level of passive stability demonstrated here already enables significant advances and applications of cryogenic microcavities without the overhead of active stabilization.

In particular, we can quantify the applicability of our passively-stabilized system to cavity quantum electrodynamics experiments with solid-state emitters, which typically require spatial tunability to locate low-loss regions of the sample or ideal emitters. When spatially and spectrally tuned to the emitter, a cavity enhances the rate of resonant emission by the so-called Purcell factor. Cavity vibrations  reduce this effect because the emitter is not always on resonance with the cavity. Assuming Gaussian statistics for cavity length excursions, the Purcell factor in the presence of vibrations is given by $F_p +1$ with
\begin{equation}
F_p(\Delta L)=F_p^0 \sqrt{\frac{\pi}{8}}\frac{\Delta L}{\sigma}e^{\Delta L^2 /8\sigma^2} erf\left(\frac{\Delta L}{2\sqrt{2}\sigma}\right),
\end{equation}
where $F_p^0 + 1$ is the resonant Purcell factor, $\Delta L$ is the cavity linewidth in length, and $\sigma$ is the rms cavity-length excursion. For example, for a cavity with a finesse $\lambda/(2\Delta L) = 16,000$ at $\lambda = 637$ nm, and $\sigma =$ 16 pm-rms as observed during quiet periods of our system, only a moderate reduction in Purcell enhancement occurs, $F_p/F_p^0 = 0.5$. 

This level of stability is particularly promising for the long-standing challenge of coupling microcavities to NV centers in diamond. Using realistic parameters from recent papers, we consider a cavity with a finesse of 11,000 in a diamond-like mode,~\cite{hoyjensenCavityEnhancedPhotonEmission2020} and a fiber mirror radius of curvature of 16 $\upmu$m.~\cite{riedelEngineeringPhotonicEnvironment2017} We assume an NV center with ideal orientation, location and quantum efficiency, a membrane of thickness 3.8 $\upmu$m,~\cite{rufOpticallyCoherentNitrogenVacancy2019} and an air gap of ${\sim} 2~\upmu$m.~\cite{riedelDeterministicEnhancementCoherent2017} With these assumptions and the 16 pm-rms vibrations we observe here, we predict a Purcell enhancement of more than 250, which, combined with a ZPL fraction of $3\%$ and a linewidth of 86 MHz (the mean value seen in~\cite{rufOpticallyCoherentNitrogenVacancy2019}), corresponds to a cooperativity just over 1. In particular, the residual vibrations of this hypothetical system reduce its performance by less than 40$\%$. Thus, even without active stabilization, it should be possible to observe significant Purcell enhancement, vastly speeding the efficiency of existing probabilistic entanglement distribution methods.~\cite{humphreysDeterministicDeliveryRemote2018} With future improvements, such stable microcavity systems could even push towards the high-cooperativity regime,  offering a platform for near-deterministic quantum network protocols.~\cite{borregaardQuantumNetworksDeterministic2019} 


\section{Acknowledgements}
The authors thank Zeno Schumacher, Huub Janssen, and Bart van Bree for insightful conversations and Yi He for laser ablation of the fiber mirror. This work was supported by the National Science and Engineering Research Council (NSERC RGPIN 435554-13,  RGPIN-2020-04095, and RTI-2016-00089), Canada Research Chairs (229003 and 231949), Fonds de Recherche - Nature et Technologies (FRQNT PR-253399), the Canada Foundation for Innovation (Innovation Fund 2015 project 33488 and LOF/CRC 229003), and l'Institut Transdisciplinaire d'Information Quantique (INTRIQ). Y. Fontana acknowledges support by a Swiss National Science Foundation fellowship. L. Childress is a CIFAR fellow in the Quantum Information Science program.  

\section{Data availability}
The data that support the findings of this study are available from the corresponding author upon reasonable request. 

\bibliography{mergedbib}

\end{document}